\documentclass{elsart}

\usepackage[square]{natbib}
\RequirePackage{xspace}
\def\kaos{{\sc Kaos}$/\!${\small A1}\xspace}

\usepackage{graphicx}
\graphicspath{{figures/}}

\usepackage{amssymb}
\journal{Nucl. Instr. and Meth. in Phys. Res. A}

\begin{document}

\begin{frontmatter}

\title{In-beam tests of scintillating fibre detectors at MAMI and at
  GSI}

\author{P.~Achenbach\corauthref{cor}},
\ead{patrick@kph.uni-mainz.de}
\author{C.~{Ayerbe Gayoso}\thanksref{PhD}},
\author{J. C.~Bernauer},
\author{R.~B\"ohm},
\author{M. O.~Distler},
\author{L.~Doria},
\author{M.~{G\'omez Rodr\'iguez de la Paz}},
\author{H.~Merkel},
\author{U.~M\"uller},
\author{L.~Nungesser},
\author{J.~Pochodzalla},
\author{S.~{S\'anchez Majos}},
\author{B. S.~Schlimme},
\author{Th.~Walcher}, \and
\author{M.~Weinriefer}
\address{Institut f\"ur Kernphysik, Johannes Gutenberg-Universit\"at,
  Mainz, Germany}

\author{L.~Debenjak},
\author{M.~Potokar}, and
\author{S.~\v Sirca}
\address{University of Ljubljana and Jo\v zef Stefan Institute,
  Ljubljana, Slovenia}

\author{M.~Kavatsyuk},
\author{O.~Lepyoshkina},
\author{S.~Minami}, 
\author{D.~Nakajima}, 
\author{C.~Rappold}, 
\author{T.R.~Saito},
\author{D.~Schardt}, and
\author{M.~Tr\"ager}
\address{GSI, Darmstadt, Germany}

\author{H.~Iwase}
\address{KEK, Japan}

\author{S.~Ajimura},
\author{A.~Sakaguchi}
\address{Graduate School of Science, Osaka University, Japan}

\author{Y.~Mizoi} 
\address{Division of Electronics and Applied Physics, Osaka
  Electro-Communication University, Japan}

\thanks[PhD]{Part of doctoral thesis.}
\corauth[cor]{Tel.: +49-6131-3925831; fax: +49-6131-3922964.}

\begin{abstract}
  The performance of scintillating fibre detectors was studied with
  electrons at the spectrometer facility of the Mainz microtron MAMI,
  as well as in a $^{12}${C} beam of 2\,$A$\/GeV energy and in a beam
  of different particle species at GSI.  Multi-anode photomultipliers
  were used to read out one or more bundles of 128 fibres each in
  different geometries. For electrons a time resolution of FWHM $\sim$
  1\,ns was measured in a single detector plane with a detection
  efficiency $\epsilon > 99\,\%$.  A time resolution of 310\,ps (FWHM)
  between two planes of fibres was achieved for carbon ions, leading
  to a FWHM $\sim$ 220\,ps for a single detector. The hit position
  residual was measured with a width of FWHM $=$ 0.27\,mm. The
  variation in the measured energy deposition was $\Delta E/E=$
  15--20\,\% (FWHM) for carbon ions.  In addition, the energy response
  to p$/\pi^+\!/$d particles was studied.  Based on the good detector
  performance fibre hodoscopes will be constructed for the \kaos
  spectrometer at MAMI and for the HypHI experiment at GSI.
\end{abstract}


\end{frontmatter}

\section{Introduction}
Detectors comprised of scintillating fibres that are packed together to
form arrays and read out via multi-channel photo-detectors have been
in use since the 1990s to track charged particles,
e.g.~\cite{Baer1994,Bosi1996,Horikawa2004}.

At the Institut f\"ur Kernphysik in Mainz, Germany, the microtron MAMI
has been upgraded to 1.5\,GeV electron beam energy and can now be used
to study strange hadronic systems. During the last years the \kaos
spectrometer was installed with large angular acceptance under forward
angles at the existing spectrometer facility for strangeness reaction
spectroscopy. A large fibre detector set-up is under development for
the \kaos spectrometer~\cite{Achenbach-GSISciRep06,
  Achenbach-GSISciRep07, Achenbach-HYP06,Achenbach-SNIC06}: the
coordinate detector of the spectrometer's electron arm will consist of
two vertical planes of fibre arrays, covering an active area of $L
\times H\sim$1600 $\times$ 300\,mm$^2$, supplemented by one or more
horizontal planes.

The HypHI project is dedicated to hypernuclear spectroscopy with
stable heavy ion beams and rare isotope beams at GSI, Germany, and
FAIR, the Facility for Antiproton and Ion Research~\cite{Saito-HYP06}.
One of the central parts of the HypHI experiment is a scintillating
fibre detector for the measurement of decay vertices of hypernuclei
and for the tracking of charged particles directly behind the target.
The fibre detector will also become crucial in distinguishing the
hypernuclei $^4_\Lambda$H and $^3_\Lambda$H from the background
containing $\alpha$ and $\Lambda$ particles produced at the target.
This detector is currently under
development~\cite{Lepyoshkina-GSISciRep06, Nakajima-GSISciRep07}.

In this paper results from beam-tests of scintillating fibre detectors
are discussed. Section~2 describes the detectors and the read-out
electronics.  In Section~3 a resume of an investigation with electrons
tracked through a large magnetic spectrometer at MAMI is given.  The
following Sections~3 and 4 show the results on position, time and
energy resolutions obtained with primary $^{12}${C} beams and a beam
of several particle species at GSI. Section~5 closes with the future
prospects of using this kind of fibre detectors and electronics with
the \kaos spectrometer at MAMI and within the HypHI project at GSI.

\section{Detector geometries and read-out electronics}
The fibres are of the standard (Non-S) type SCSF-78 ({\sf Kuraray},
Japan) with double cladding and $\oslash=$ 0.83\,mm outer diameter.
The cladding thickness is $d_\mathrm{clad}\approx 0.1\,$mm, leading to
a 0.73\,mm thick core made of a polystyrene base with refractive index
$n_\mathrm{core}=$ 1.6. The outer cladding is made of a fluorinated
polymer with refractive index $n_\mathrm{clad'}=$ 1.42 and the inner
cladding is made of polymethylmethacrylate with refractive index
$n_\mathrm{clad}=$ 1.49. The emission spectrum extends from
$\lambda\sim$ 415--550\,nm with a maximum at $\lambda\sim$ 440\,nm for
fibres of short lengths.  For a double cladding fibre the critical
axial angle is given by $\theta_\mathrm{crit} = \arccos
n_\mathrm{clad'}/n_\mathrm{core}= 26.7^\circ$. The trapping efficiency
for photons produced close to the axis of the fibre is $5.3\,\%$
giving $70\,\%$ more light than single cladding fibres, where the
efficiency for light trapped inside the core is only $3.1\,\%$.

The fibre arrays consisted of 4 double layers. A fibre column was
formed by 4 fibres, one from each double layer, coupled to one common
read-out channel of the photo-detector by a small plastic cookie with
a matrix of holes. These columns were aligned along the incident
particle direction. The read-out was one-sided by multi-anode
photomultipliers (MaPMTs). No optical grease was used between fibres
and the entrance window. The MaPMTs of type R7259K ({\sf Hamamatsu
  Photonics}, Japan) are fitted with a 32-channel linear array of
metal electrodes. The photocathode material is bialkali and the window
is made of 1.5\,mm thick borosilicate glass.  The effective area per
channel is $0.8 \times 7$\,mm$^2$ with a pitch of 1\,mm. A
complication arises from the fact, that the numerical aperture for the
multi-cladding fibre is 0.72, so that the light cone at the
photo-cathode has a diameter of 2.32\,mm.  This results in a hit
multiplicity of typical 3 neighbouring channels.

The MaPMTs have been characterised with an average anode luminous
sensitivity of $S_a= 374\,$A$/$lm (according to data sheet 140\,A$/$lm
is typical), an average cathode luminous sensitivity $S_c=
84\,\mu$A$/$lm (70\,$\mu$A$/$lm typical), and an average gain $G= 4.4
\cdot 10^6$ ($2\cdot 10^6$ typical). The gain uniformity between
anodes was found to be between 1:1.1 and 1:1.25 (1:1.5 typical), with
the edge anodes having slightly lower gains on average. Hardly any
strip has less than 70\% of the maximum gain of a given
photomultiplier.

Instead of supplying dynode voltages through a voltage divider the
MaPMTs were powered by individual Cockcroft-Walton bases ({\sf HVSys},
Russia). The dc voltage is pulsed and converted with a ladder network
of capacitors and diodes to higher voltages. Stiff high voltage cables
are not needed, since only $\sim$140\,V has to be delivered to the
first front-end board, where the voltage is daisy-chained to the other
boards of the detector plane.

For the construction of the fibre bundles position matrices were
designed, allowing for the mounting and the alignment of the fibres
with the desired pitch.  The construction procedure for the fibre
bundles was to glue single layers of 16 fibres into grooved aluminium
plates with acrylic white paint. A total of 8 single layers is needed
for each bundle. These bundles were then glued into the cookies. The
bundles were polished using a diamond cutting tool ({\sf Mutronic},
Germany). Two of the three bundles had to be bent.  This was done by
placing them into an oven at a temperature of 70\,$^\circ$C for about
1\,hour and bending them into the desired shape.
 
As half of the scintillation light is emitted in the direction
opposite to the photomultiplier, light reflected by the end face can
contribute significantly to the total light output.  In order to
increase the light yield, some of the fibre bundles used in the tests
were aluminised at the polished free end face for a high reflectivity.
A vaporisation chamber was utilised that consisted of a vacuum chamber
with an electric oven in which a small pellet can be placed.  The
aluminium coating adhered firmly and smoothly to the end face of the
bundle.  The increase of light yield was confirmed with a $^{90}$Sr
source in laboratory measurements and amounted to $\sim$ 50\,\% .

\begin{figure}[ht]
  \centering
  \includegraphics[width=0.58\textwidth]{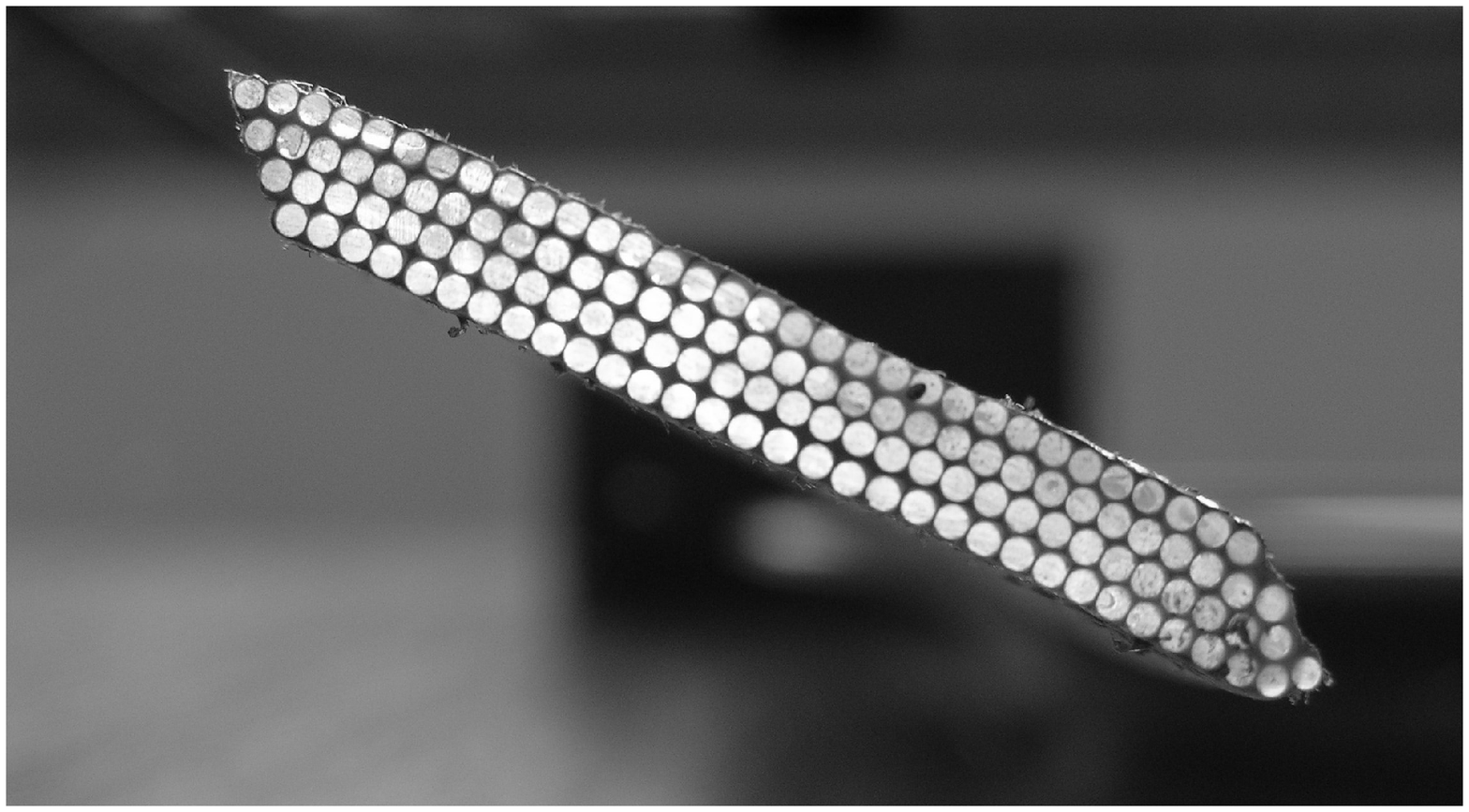}
  \includegraphics[width=0.38\textwidth]{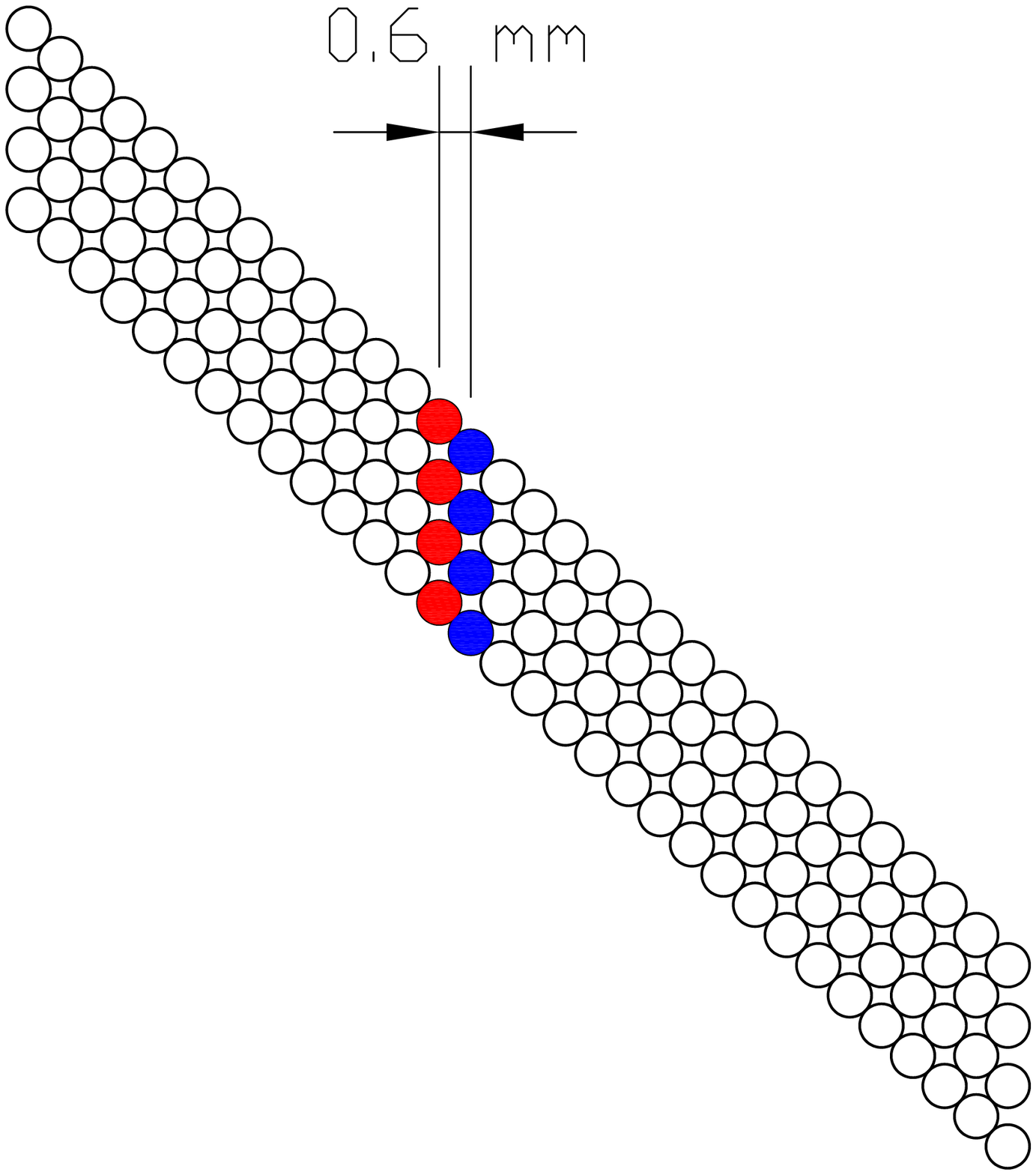}
  \caption{Scheme of a 45$^\circ$ column angle geometry with 4 double
    layers of fibres and a column pitch of 0.6\,mm and an overlap of
    0.24\,mm (right). A photograph of an assembled fibre bundle
    (left).}
  \label{fig:45deg}
\end{figure}

The beam-tests at MAMI were performed with a fibre detector positioned
near the focal plane of a large magnetic spectrometer where scattered
electrons have an inclination of 45$^\circ$ with respect to the normal
to the detection plane.  If the particles were crossing the fibre
array with an angle to the column direction its tracking capability
would be compromised. Accordingly, two bundles of 4 double layers with
the fibre columns following the 45$^\circ$ inclination in square
packing geometry were built.  The fibre array has an overlap of $o=
\oslash (1 - 1/\sqrt{2})\approx$ 0.24\,mm, and a column pitch of $p=
\oslash/\sqrt{2}\approx$ 0.59\,mm, see Fig.~\ref{fig:45deg}. Our
experience shows that the best way to stack the fibres during the
actual assembly is layer-by-layer. Other sequences may cause sizeable
misalignments which directly lead to errors in the position
determination.

\begin{figure}[ht]
  \centering
  \includegraphics[width=0.55\textwidth]{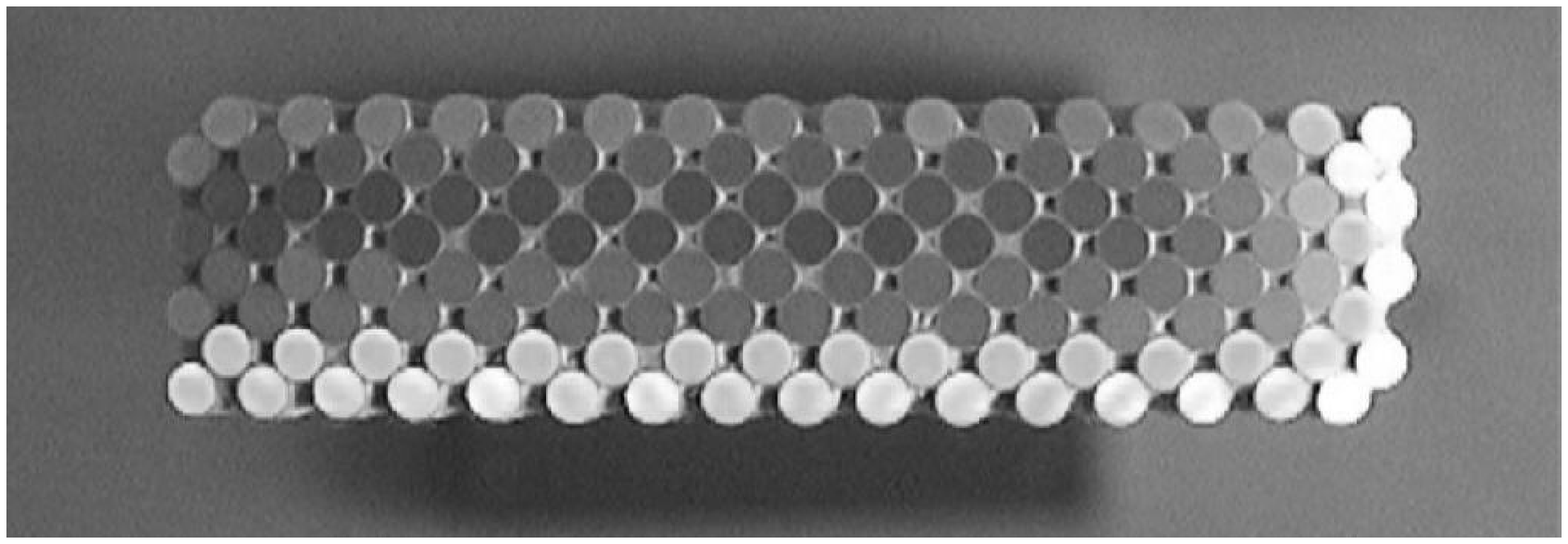}
  \includegraphics[width=0.43\textwidth]{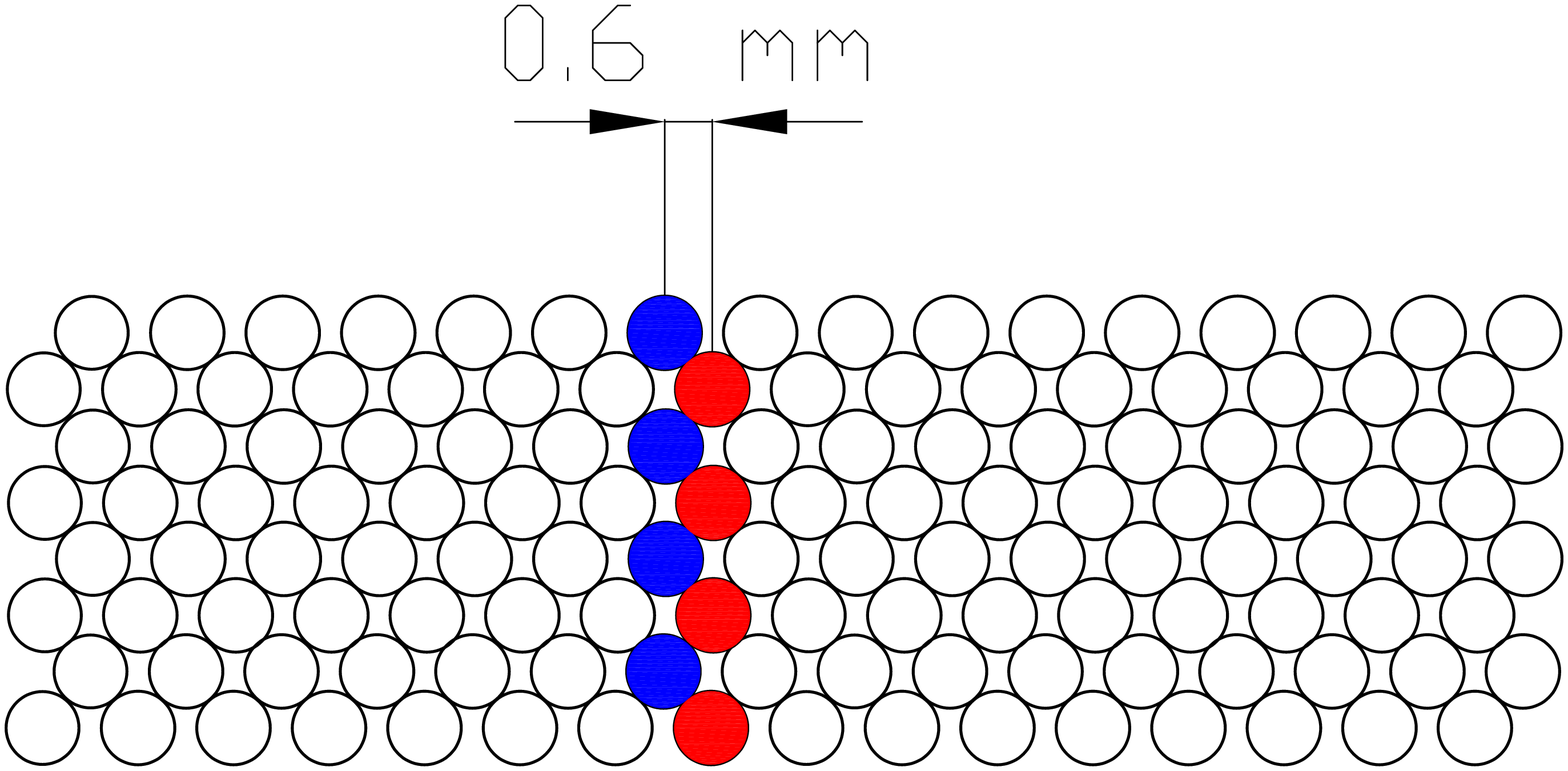}
  \caption{Scheme of a 0$^\circ$ column angle geometry with 4 double
    layers of fibres and a column pitch of 0.6\,mm and an overlap of
    0.24\,mm (right). A photograph of an assembled fibre bundle
    (left).}
  \label{fig:0deg}
\end{figure}

For the tests in the carbon beam 3 bundles of 128 fibres each with an
active area of $150 \times 20$\,mm$^2$ were constructed in a square
packing geometry of 4 double layers, see Fig.~\ref{fig:0deg}. Overlap
and column pitch of this geometry are identical to the 45$^\circ$
geometry. Results from for laboratory measurements with a $^{90}${Sr}
source resulted in a light yield of 4--5 photoelectrons per pixel
with a multiplicity of 3 pixels, corresponding to 15 photoelectrons
per crossing minimum ionising particle.

\begin{figure}[ht]
  \centering
  \includegraphics[width=0.6\textwidth]{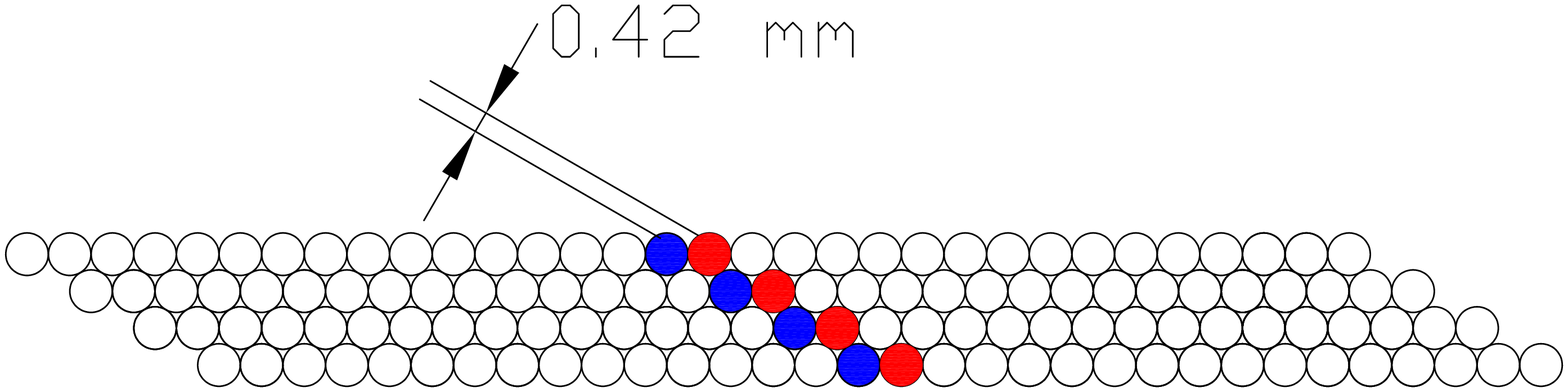}\\[2mm]
  \includegraphics[width=0.6\textwidth]{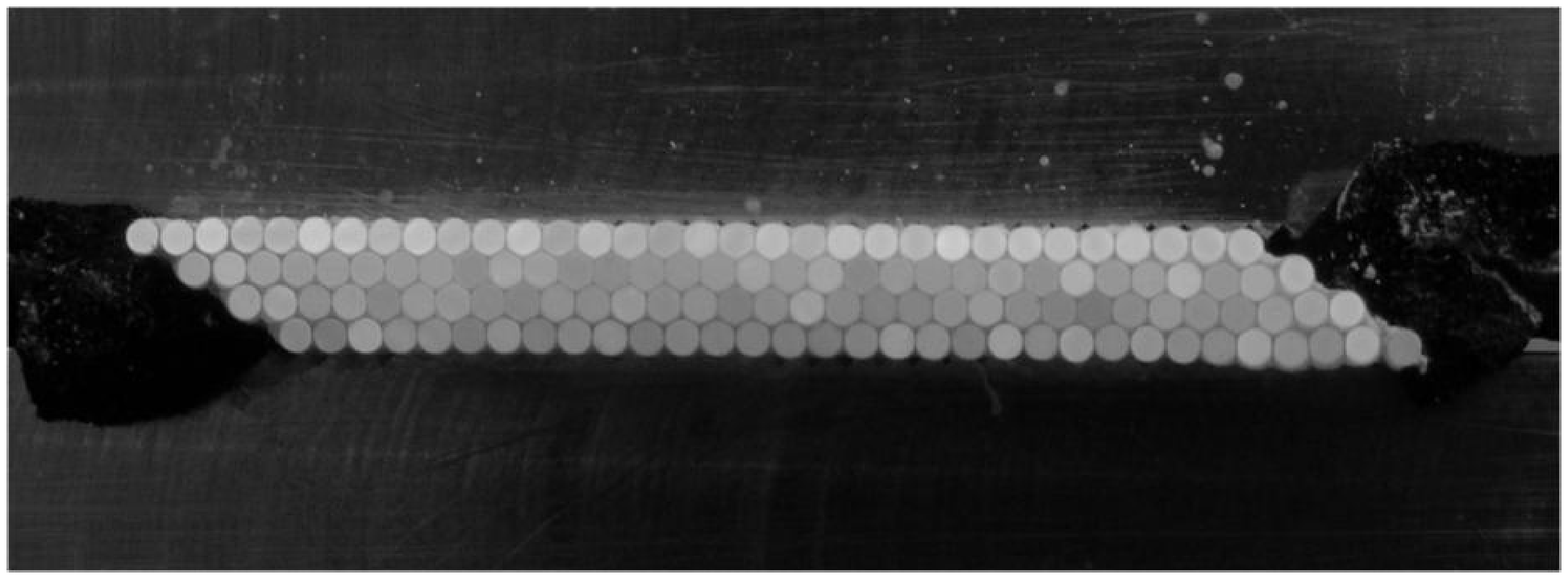}
  \caption{Scheme of a 60$^\circ$ column angle geometry with 4 double
    layers of fibres and a column pitch of 0.42\,mm and an overlap of
    0.42\,mm (top). A photograph of an assembled fibre bundle
    (bottom).}
  \label{fig:60deg}
\end{figure}

In the \kaos spectrometer, particles will cross the electron arm focal
plane with an inclination angle of $50-70^\circ$ with respect to the
normal of the plane. Regarding this geometry fibre arrays with a
hexagonal packing of slanted columns with 60$^\circ$ inclination were
constructed, see Fig.~\ref{fig:60deg}. Overlap and column pitch of
this geometry are 0.41\,mm. These detectors were tested in the beams
at GSI.

A 12-layer front-end board able to accommodate the three MaPMTs was
developed by the Institut f\"ur Kernphysik for this type of fibre
detector. It supplies the voltage for the Cockcroft-Walton voltage
multipliers and brings the analogue signals along equalised conducting
paths to the RJ-45 connectors for the output to the discriminators.
For amplitude-compensated timing two 32-channel discriminator boards,
custom designed and built by the the Institut f\"ur Kernphysik, each
with 4 integrated low-walk double threshold discriminators (DTDs),
were used. The DTD boards were placed in an 6U crate together with a
controller board. The communication with a PC was done via parallel
port. The time is picked off by CATCH cards, developed for the COMPASS
collaboration~\cite{COMPASS2001}. At GSI analogue output boards with
50\,$\Omega$ coaxial connectors were attached to the discriminator
boards to access also the pulse heights of each channel. The pulse
height information of 22 channels was read out by two Model 2249A
CAMAC ADCs ({\sf LeCroy}, US). The ADCs offer a resolution of 10\,bits
at an input sensitivity of 0.25\,pC$/$count.

The trigger was either derived from the second plane of the fibre
detector or from a reference counter, a scintillation paddle that was
installed 2\,m upstream in the beam-line. The time difference between
individual channels and the reference counter showed a dependence on
the pulse height. This time walk effect was corrected according to the
equation $\Delta t_c= \Delta t - c_\mathrm{walk}/\sqrt{Q}$, where
$c_\mathrm{walk}$ is a global correction coefficient and $Q$ the
charge of the signal as measured by the ADC. Any time jitter from the
reference counter drops out when the residual time between two
detector planes is determined.

\section{Performance of a fibre detector with electrons at the Mainz
  microtron MAMI}
In order to measure the tracks of electrons through a 32-channel fibre
detector, it was placed inside the heavy shielding house of
spectrometer~A, a large high-resolution magnetic spectrometer at the
Mainz Microtron MAMI~\cite{Blomqvist1998}. The focal plane of the
spectrometer~A has a length of approximately 2\,m, and it is inclined
at an angle of 45$^\circ$ to the reference particle trajectory. The
divergence of the particle trajectories is about 24$^\circ$. The fibre
detector was sandwiched between the drift chambers and the
scintillator paddles of the focal plane detector system.

\begin{figure}[ht]
  \centering
  \includegraphics[width=0.49\textwidth]{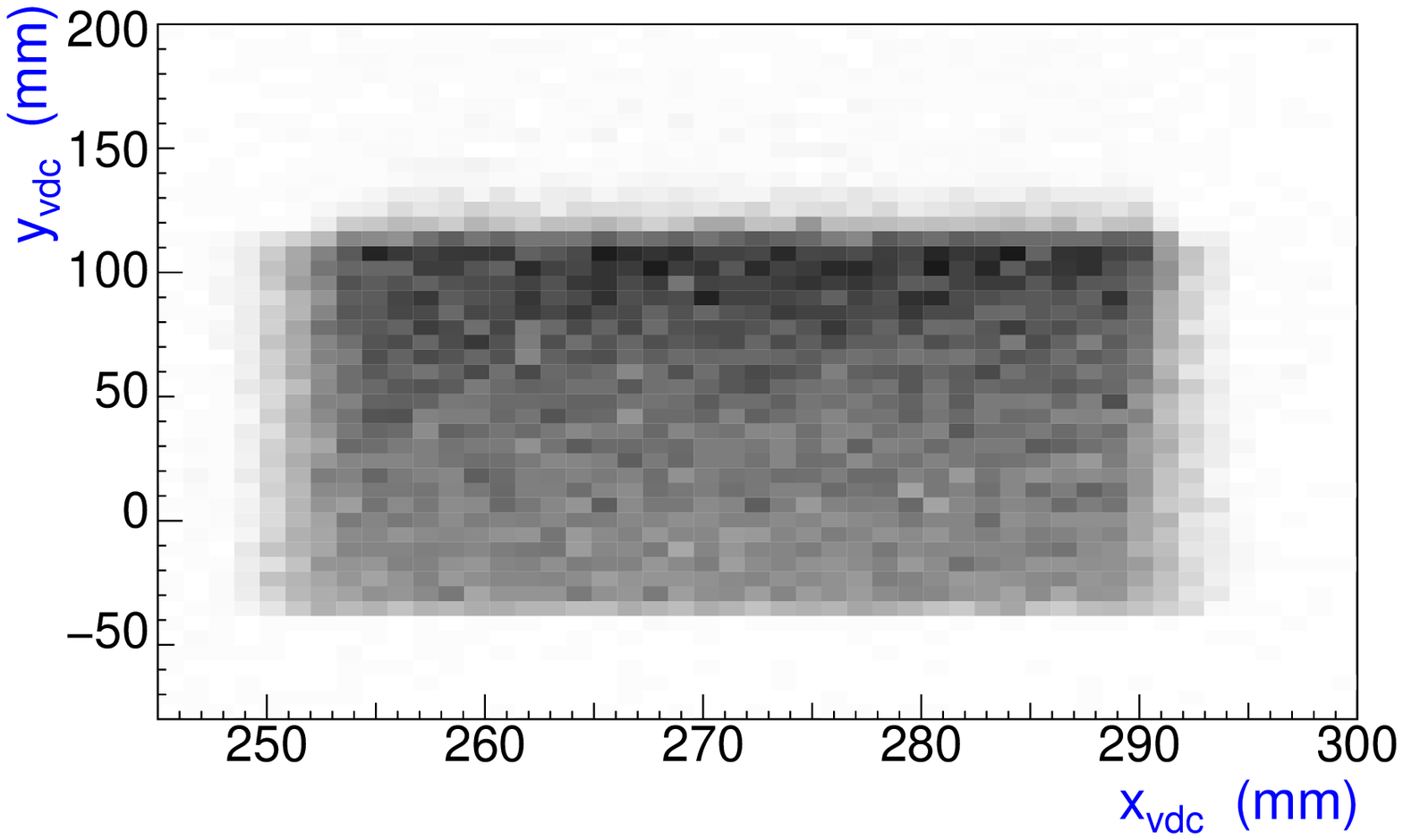}\hfill
  \includegraphics[width=0.49\textwidth]{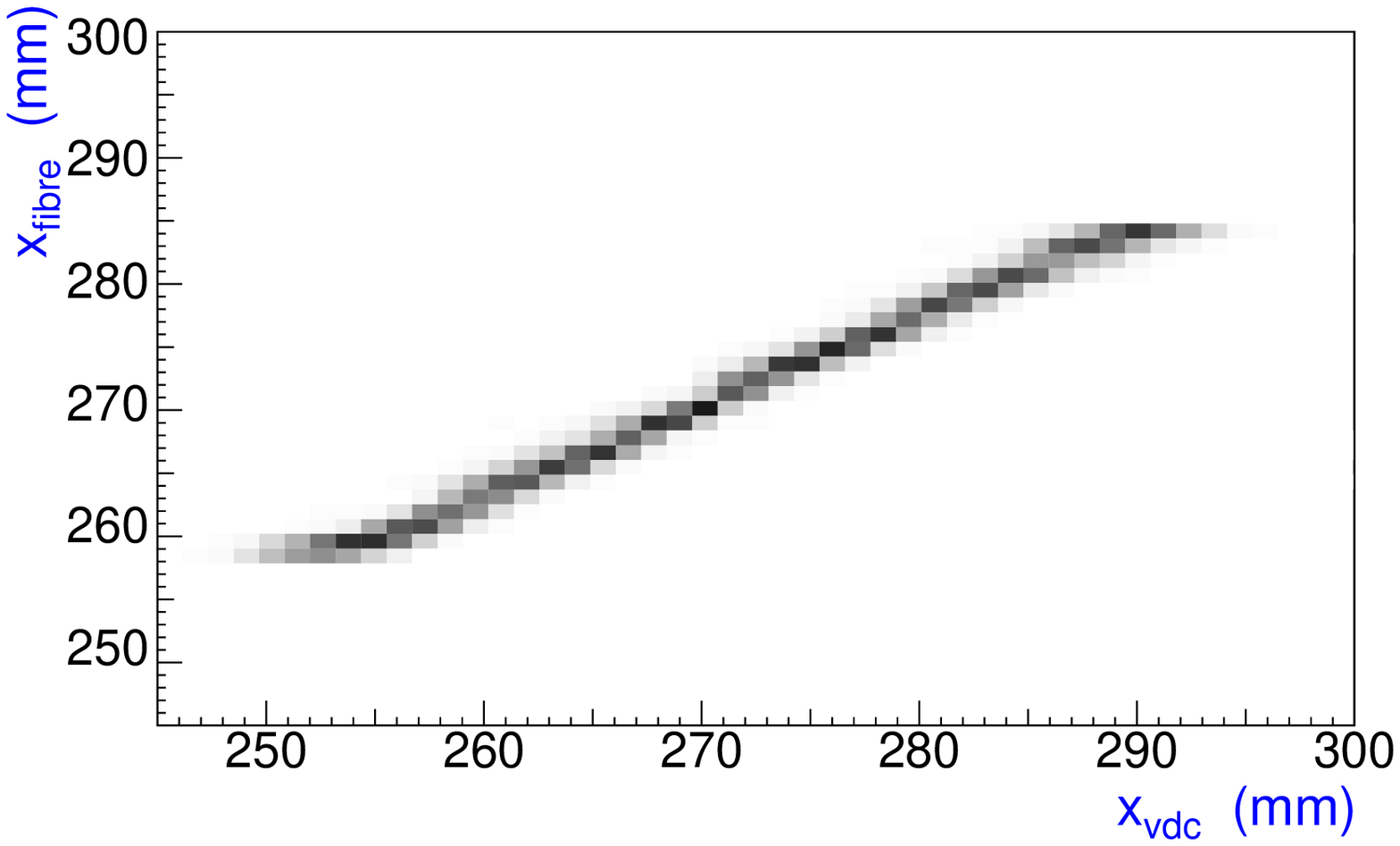}
  \caption{Geometrical acceptance covered by the fibre detector inside
    the spectrometer shown by the track positions reconstructed with
    the vertical drift chambers (left).  The reconstructed position
    projected onto the base coordinate versus the measured position
    obtained from the fibre detector with a simple estimator (right).}
  \label{fig:MAMI-VDC}
\end{figure}

Two vertical drift chambers (VDCs) measure the dispersive coordinate
$x$ plus the corresponding angle $\theta$ and the non-dispersive
coordinate $y$ plus the angle $\phi$. The electron hit position was
found by extrapolating the reconstructed track from the focal plane to
the fibre detector. Fig.~\ref{fig:MAMI-VDC}\,(left) shows the
geometrical acceptance covered by the fibre detector inside the
spectrometer, determined by such an extrapolation. The kinematics of
the reaction was chosen so that the particle illumination was
homogeneous over the fibre detector.

\begin{figure}[ht]
  \centering
  \includegraphics[width=0.49\textwidth]{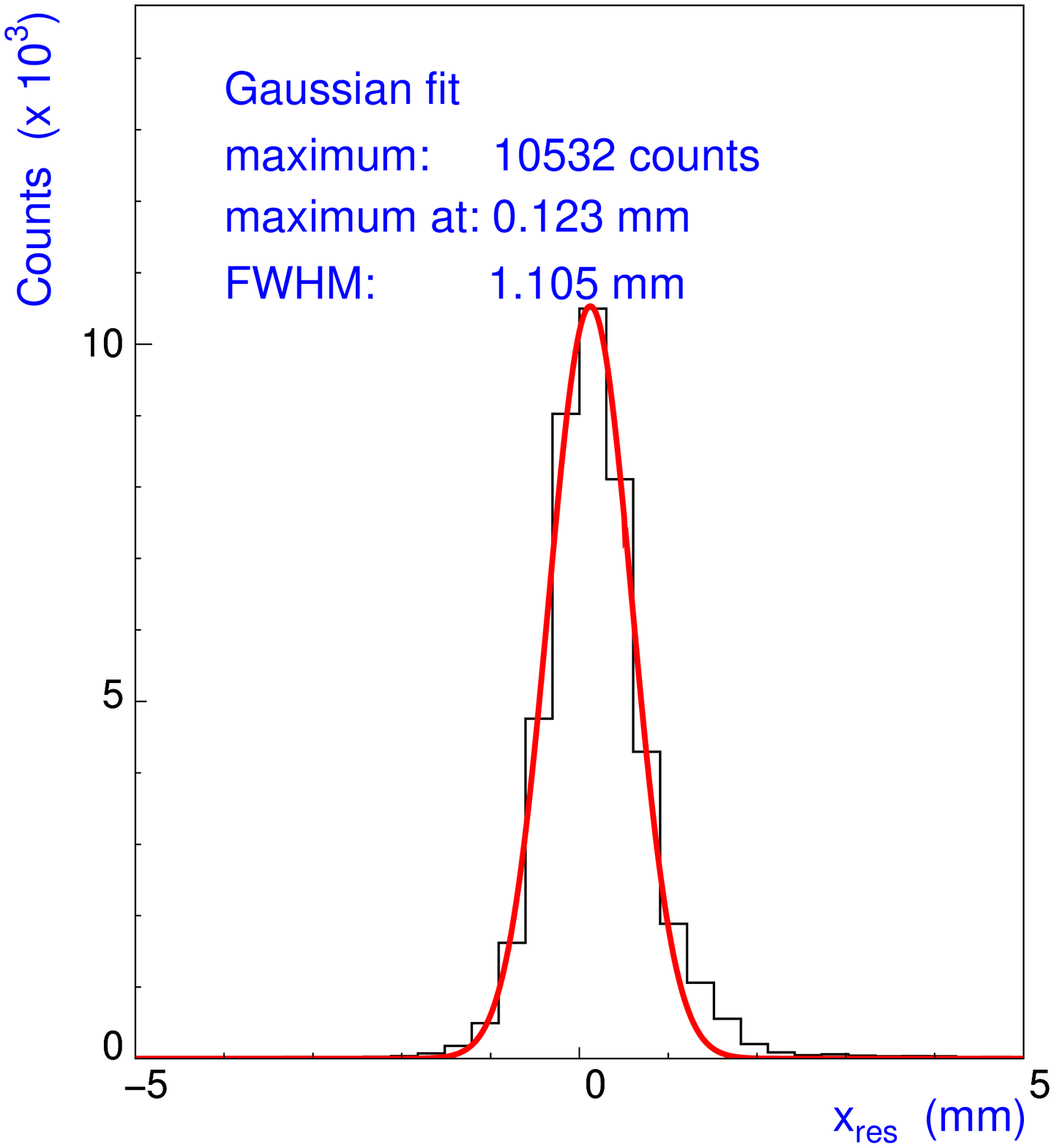}\hfill
  \includegraphics[width=0.49\textwidth]{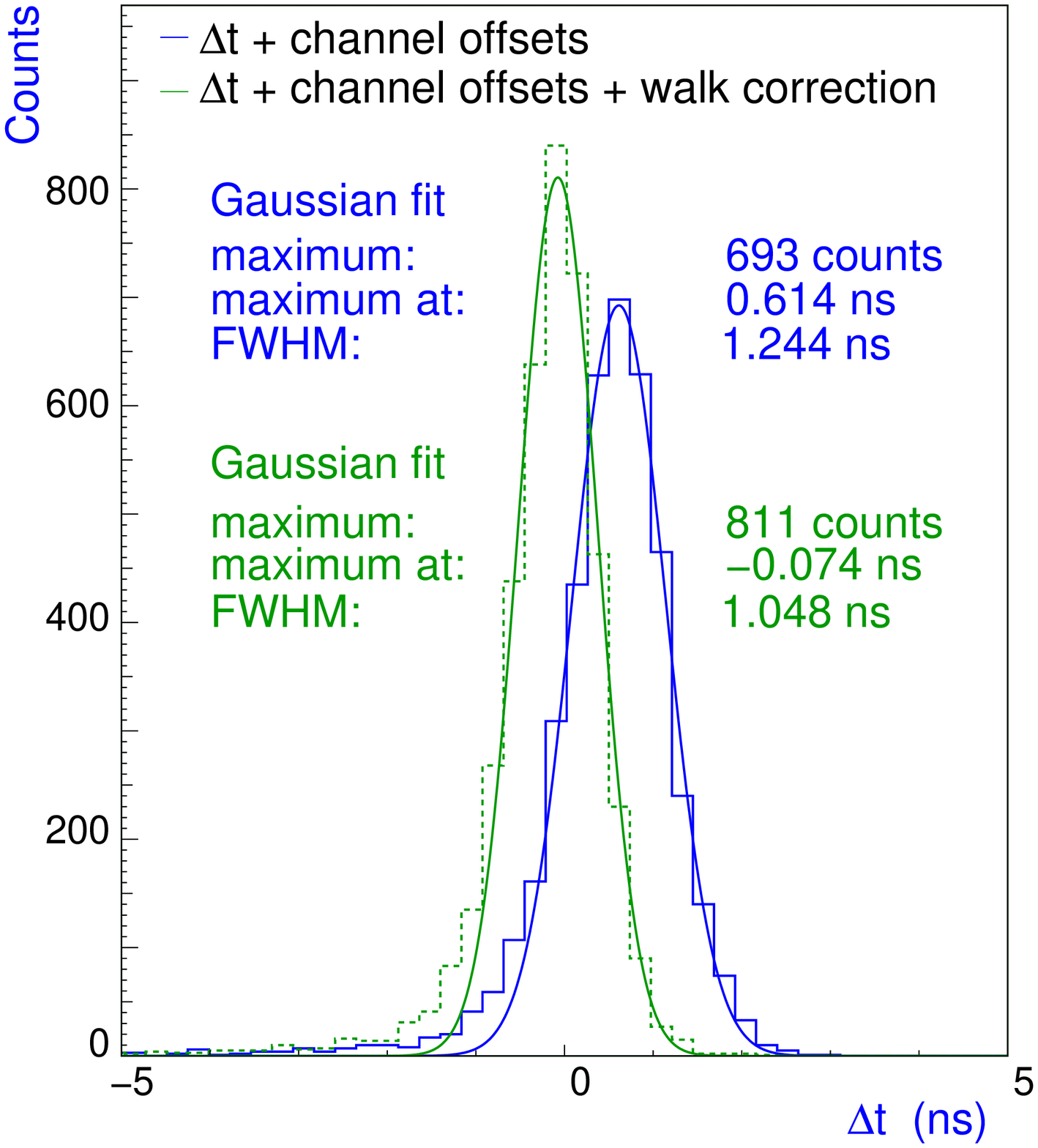}
  \caption{The residual of track position obtained from the drift
    chamber track reconstruction and the simple estimator for the
    fibre detector (left). The residual of hit times obtained from the
    trigger scintillators and the fibre detector before and after walk
    correction for the trigger and calibration of the fibre
    channel-to-channel variations (right).}
  \label{fig:MAMI-residua}
\end{figure}

The hit multiplicity of the detector was relatively high, $N\approx
4$, the main reason being a large optical cross-talk in the MaPMT. A
simple estimator for the $x$-position of the form $x= \sum_{i=1}^{N}
x_i/N$ was used accordingly, where $x_i$ was the parametrised
geometrical centre position of the $i$th channel and $N$ the hit
multiplicity.  This track estimate was compared to the track
reconstructed from the VDCs and projected onto the detector base
coordinate, see Fig.~\ref{fig:MAMI-VDC}\,(right). Small
non-linearities at the edges of the diagonal line indicate that better
estimators based on weighted averages were needed. A detailed analysis
of the correlation revealed a few misalignments in the fibre bundles
caused by the demanding construction of the $45^\circ$ geometry.
Fig.~\ref{fig:MAMI-residua}\,(left) shows the residual of track
position defined as the difference between the position reconstructed
by the VDCs and the position measured by the fibre detector. A width
of FWHM $\sim$ 1.1\,mm can be deduced from the distribution.  It is
assumed that the resolution of the VDC, $\Delta x <$ 100\,$\mu$m for
the dispersive coordinate, was high compared to the fibre detector.
%Because of the large multiplicities and the simple estimator the
%achieved position resolution was only mediocre, so that these aspects
%were much improved in later measurements.

The trigger detectors of the spectrometer consist of two segmented
planes of plastic scintillation detectors. The arrival time of the
electrons was measured in the fibre detector with respect to the
following two overlapping paddles.
Fig.~\ref{fig:MAMI-residua}\,(right) shows the time spectra obtained
from the coincidence timing with the trigger scintillators before and
after performing the walk correction for the paddles and the
calibration of the channel-to-channel variations of the fibre
detector. The combined resolution of FWHM $\approx 1$\,ns was rather
good for the small amount of light from the fibres.

The detection efficiency was determined by sandwiching the fibre
detector between the VDC and the scintillators, and using the
three-detector method. It was found to be 99\,\% independent of the
threshold.

\section{\boldmath Performance of a fibre detector in a carbon
  at GSI}
In Cave~C of GSI tests of a fibre detector with three bundles in a
$^{12}${C} beam of 2\,$A$\/GeV energy were performed.  Two bundles of
the detector were aligned to a single plane, and one bundle formed a
parallel plane directly behind.

\begin{figure}
  \centering
  \includegraphics[width=0.6\textwidth]{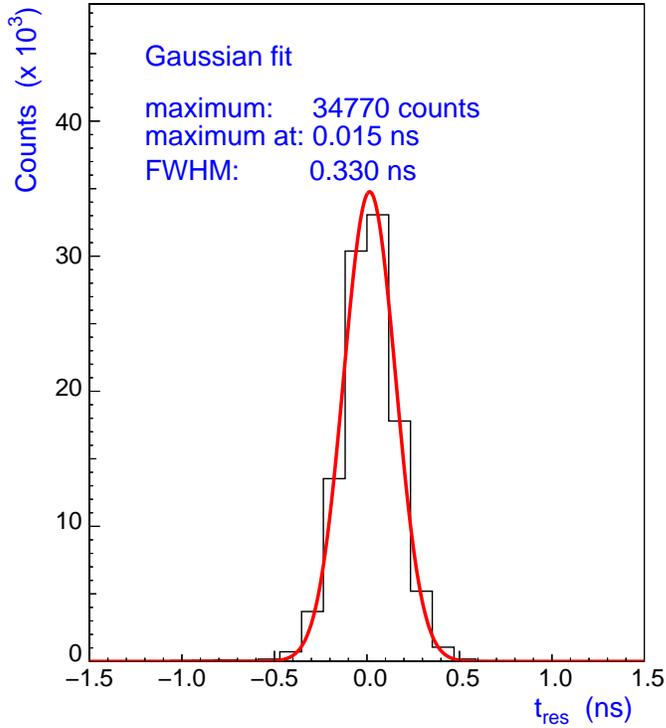}
  \caption{The residual of hit times between two detector planes, $t_A
    - t_B$. A Gaussian fit is shown providing a width of FWHM $=$
    330\,ps. The time resolution of a single detector plane was
    derived to be FWHM $\sim$ 330\,ps$/\sqrt{2}=$ 230\,ps.}
  \label{fig:GSIC12-tresidual}
\end{figure}

In deducing the time resolution, an iteration over all hits in a plane
including multiple hits in a channel was performed, and clusters of
correlated hit times were searched for. The cluster with the time
closest to the trigger signal time was retained, and within the
cluster the time of the first arrived signal was chosen as hit time.
In the algorithm a minimum time separation of 10\,ns between clusters
and a hit in a coincidence window of 20\,ns width were required. A
time walk correction for the hit times was not needed.  The hit time
residual, defined as the difference between the two hit times in the
two planes of fibres, was distributed with a width of FWHM $=$ 330\,ps
for the carbon beam, see Fig.~\ref{fig:GSIC12-tresidual}.  No
significant dependence of the time resolution on photomultiplier high
voltage was observed.  The time resolution of a single detector plane
was derived to be FWHM $\sim$ 330\,ps$/\sqrt{2}=$ 230\,ps.

\begin{figure}[ht]
  \centering
  \includegraphics[width=0.6\textwidth]{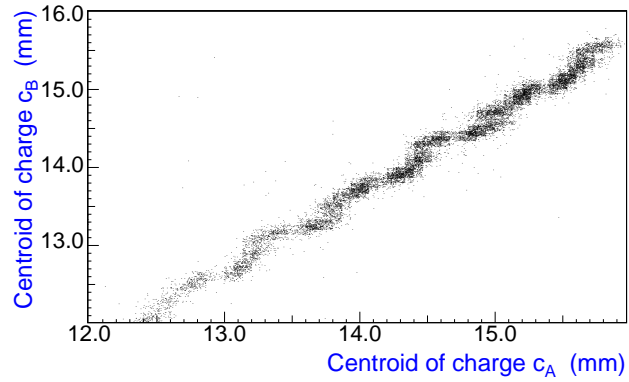}
  \caption{Scatter plot showing the correlation between the centroids
    of charges in both detector planes. The steps appear with a pitch
    of $\Delta c_A, \Delta c_B \approx$ 0.6\,mm.}
  \label{fig:GSIC12-ABposition}
\end{figure}

\begin{figure}[ht]
  \centering
  \includegraphics[width=0.6\textwidth]{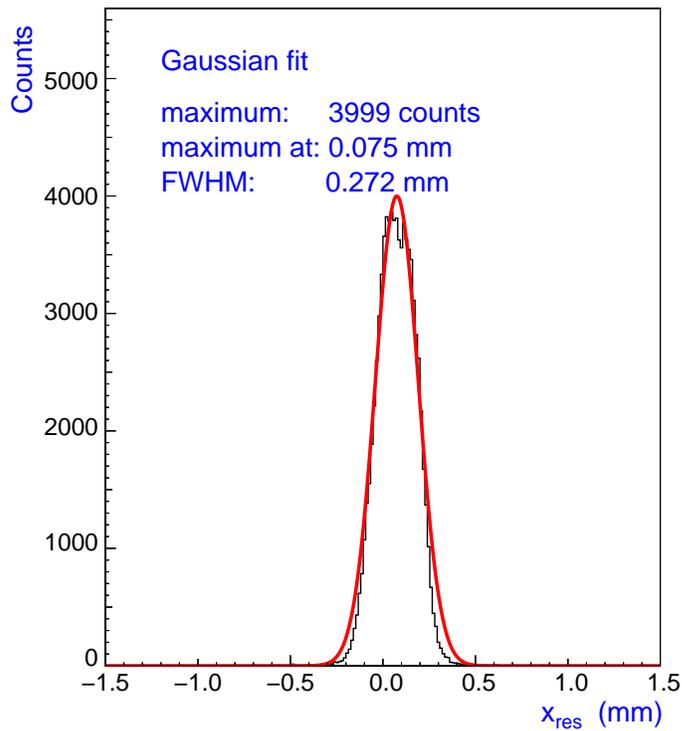}
  \caption{The residual of track position estimates between both
    detector planes, $x_A - x_B$, using the centroid of charges. A
    Gaussian fit is shown providing a width of FWHM $=$ 0.27\,mm,
    however, the distribution is non-Gaussian with two overlapping
    peaks because of the discretisation in fibre channels.}
  \label{fig:GSIC12-xresidual}
\end{figure}

The pulse height spectra for different high voltages ($-650$, $-700$,
and $-850$\,V) were fitted to get the normalisation, $A_i$, and the
pedestal position, $p_i$, for each ADC channel $i$. With these
calibration values the spectra were corrected for channel-to-channel
gain variations. For 11 neighbouring channels in one plane and the
same number of channels in the plane directly behind the pulse height
information was available, and the hit channel was determined by the
pulse height maximum. To validate the assignment the centroid of
charges of up to five channels was determined.
Fig.~\ref{fig:GSIC12-ABposition} shows the correlation between the
centroids of charges in both detector planes.  The steps are a
consequence of the discretisation in fibre channels and appear with a
pitch of $\Delta c_A, \Delta c_B \approx$ 0.6\,mm. Taking the centroid
of only a limited number of hit channels proved to be superior to the
centroid of all hit channels. The algorithm was applied to the central
area of the detector away from the edges, where the number of channels
available for the averaging is restricted. The hit position residual,
defined as the difference between the two track position estimates in
the two planes of fibres, was measured with a FWHM $\sim$ 0.27\,mm,
see Fig.~\ref{fig:GSIC12-xresidual}.  This accuracy was sufficient for
an unambiguous identification of the hit channel, leading to a spatial
accuracy of 0.6\,mm$/\sqrt{12}\approx$ 170\,$\mu$m (rms). In
principle, the resolution can be improved by distinguishing between
hits in the overlap region of two neighbouring channels from central
hits. An analysis, which required a minimum charge drop of 10\,\%
compared to the neighbours was performed. The identification of the
hit channel was then improved and the identification of overlapping
channels became possible, but a large optical cross-talk interferes
with such a requirement. The channel with the pulse height maximum was
strongly correlated to the hit time defining channel showing that ADC
calibration constants and TDC off-set values had both been correctly
determined.

It is worth estimating the spatial accuracy of the fibre detector for
set-ups without analogue read-out. Cross-talk between neighbouring
channels then perturbs the reconstruction of the position of the track
position by causing finite hit multiplicity. Since no absolute
position of the particle tracks were known, one has to compare the hit
channel mean value with the estimated position using the ADC
information which is assumed to be more accurate.  The difference of
the hit channel mean value as a simple estimate of track position and
the reconstructed position was evaluated by requiring a certain
minimum ADC value (typically 70\% of the maximum) to mimic a given
discriminator voltage threshold.  This difference still includes
contributions from the uncertainties in both position estimators and
the granularity of the fibre array.  The distributions were measured
with an average $\langle$FWHM$\rangle$ $\sim$ 0.6\,mm (approx.\ the
fibre pitch), an average $\langle$RMS$\rangle$ of 0.5\,mm, and an
average $\langle$FWTM$\rangle$ $<$ 1\,mm.

\begin{figure}[ht]
  \centering
  \includegraphics[width=0.48\textwidth]{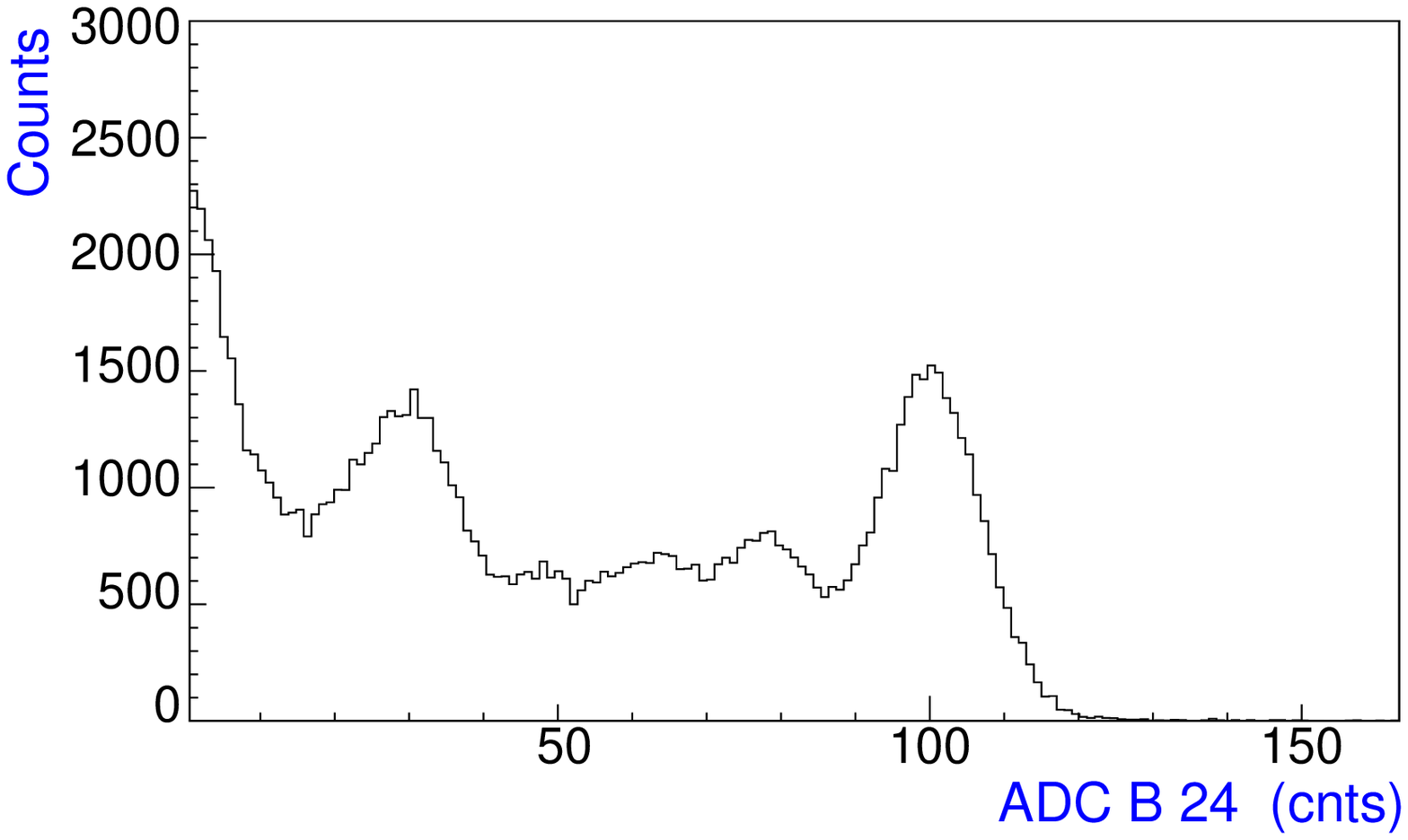}
  \includegraphics[width=0.48\textwidth]{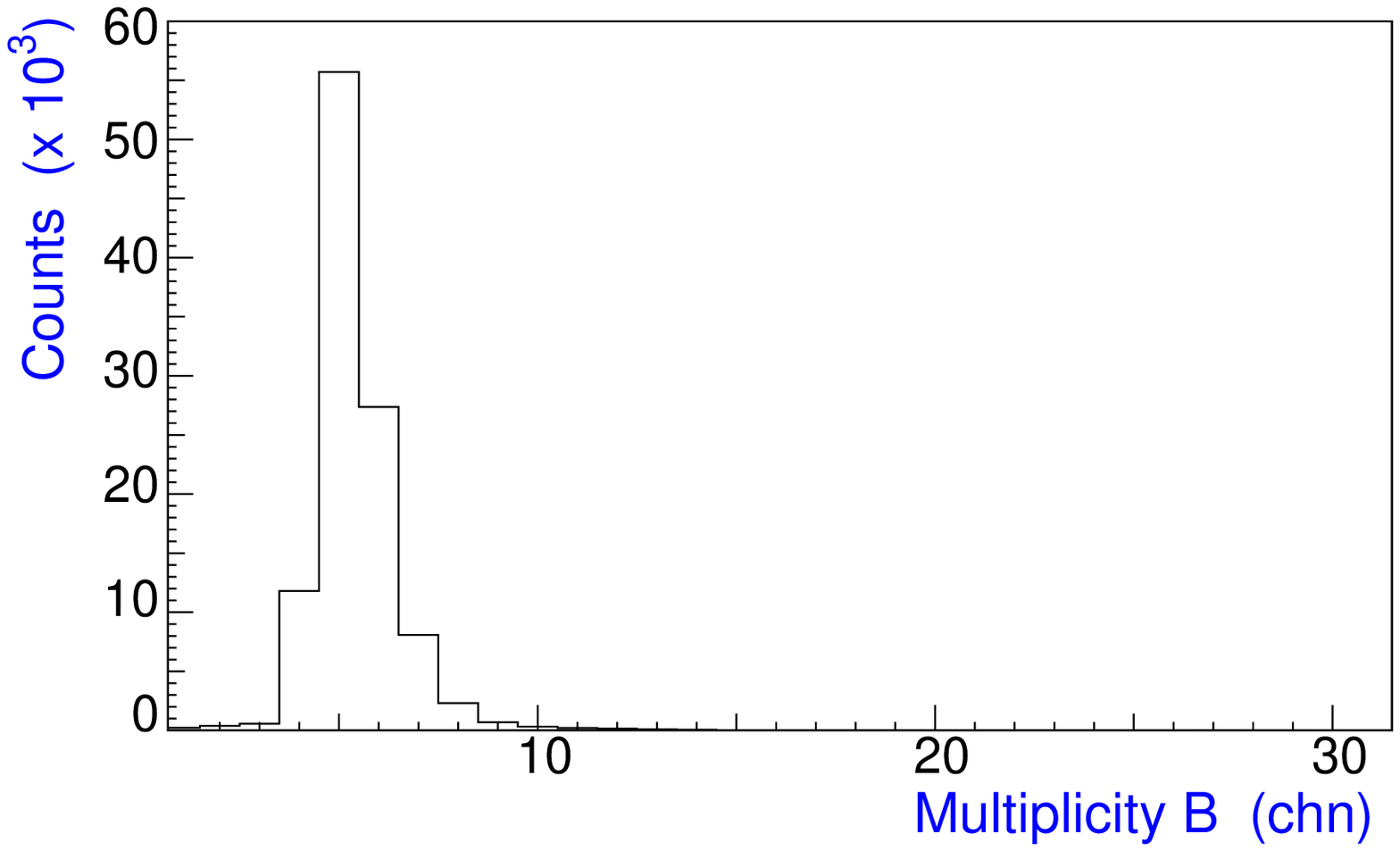}
  \caption{Pulse height distribution of a typical detector channel
    (left). Distribution of hit multiplicities in the corresponding
    detector plane (right), which causes the series of peaks in the
    pulse height distribution.}
  \label{fig:GSIC12-pulseheight}
\end{figure}

The pulse height distribution of a typical detector channel (B~24) is
shown in Fig.~\ref{fig:GSIC12-pulseheight}\,(left). The appearance of
a series of peaks below the maximum pulse height at ADC channel $\sim$
100 was caused by the spread of secondary electrons and the cross-talk
between channels that transport a fixed fraction of the signal into
neighbouring channels. The hit multiplicity of one detector plane is
shown in Fig.~\ref{fig:GSIC12-pulseheight}\,(right). For a high
voltage of $-650$\,V the mean value of the distribution was $N\sim$ 5,
increasing with higher voltages. The overall photomultiplier gain
increased by a factor $\sim$ 2 between $-650$\,V and $-850$\,V. From
the distributions of the pulse height sum over all channels a relative
energy resolution of $\Delta E/E=$ 15--20\,\% was determined. The
detection efficiency of a plane, i.e.\ the probability to find at
least one hit in one plane provided a hit in the other plane, was
above 99\,\%.

\section{\boldmath Performance of a fibre detector in a beam of
  different particle species at GSI}
A fibre detector in Cave~A of GSI was tested in a p$/\pi^+\!/$d beam
of 3.3\,Tm magnetic rigidity with dominant protons of 1\,GeV$/$c
momentum as well as in a carbon beam of 2\,$A$\/GeV energy.
 
\begin{figure}[ht]
  \centering
  \includegraphics[width=0.6\textwidth]{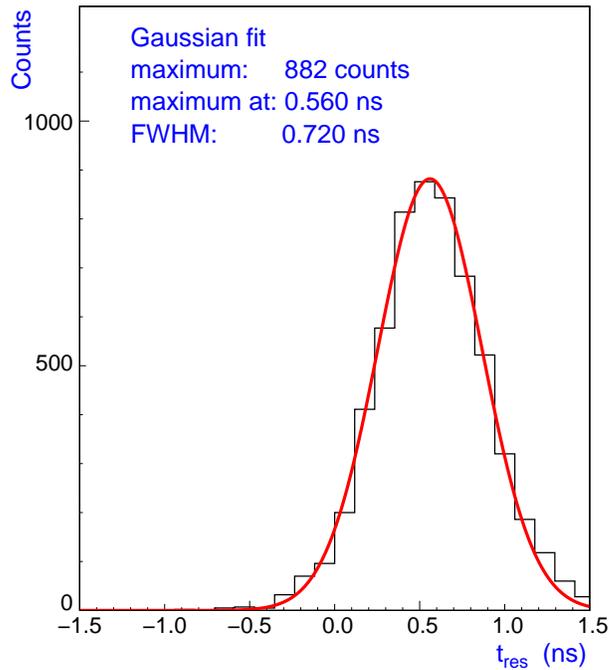}
  \caption{The residual of hit times between two detector planes, $t_A
    - t_B$. A Gaussian fit is shown providing a width of FWHM $=$
    720\,ps for the beam of different particle species.  The time
    resolution of a single detector plane was derived to be FWHM
    $\sim$ 510\,ps.}
  \label{fig:GSIcb-piontiming}
\end{figure}

The hit time residual was measured with a width of 720\,ps for the
beam of different particle species, see
Fig.~\ref{fig:GSIcb-piontiming}.  The time resolution of a single
detector plane was derived to be FWHM $\sim$ 310\,ps$/\sqrt{2}=$
220\,ps for the carbon beam and 510\,ps for the beam of different
particle species.  For the measurements in the carbon beam the high
voltages were reduced from $-$850\,V to $-$600\,V to compensate for
the large pulse heights due to the large energy deposit. Of course,
statistical fluctuations in the number of detected photons were not
affected by this reduction of the gain.

\begin{figure}[ht]
  \centering
  \includegraphics[width=0.6\textwidth]{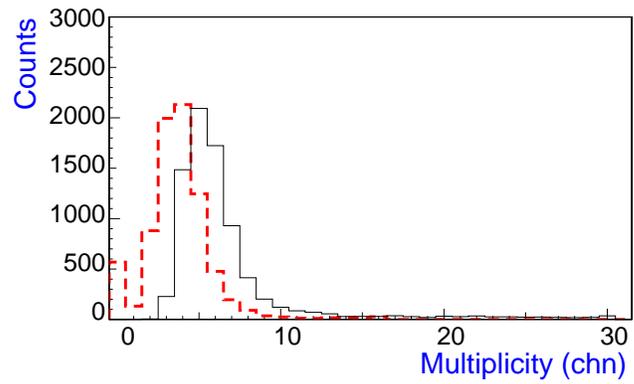}
  \caption{The hit multiplicities of both detector planes for the beam
    of different particle species. The trigger was provided by the
    plane represented with the dashed curve.}
  \label{fig:GSIcb-multi}
\end{figure}

The multiplicity distributions of both detector planes and are shown
in Fig.~\ref{fig:GSIcb-multi} with average values close to $N=$ 5
channels.  These values are the consequence of some small misalignment
and mainly cross-talk in the glass window of the MaPMT.  By using the
pulse height information the hit channel was determined as the
centroid of charges.  The hit position residual, defined as the
difference between the two estimates in the two planes of fibres, was
fitted with a width of FWHM $=$ 0.46\,mm for the beam of different
particle species. It was to some extent compromised by gain
variations.

\begin{figure}[ht]
  \centering
  \includegraphics[width=0.6\textwidth]{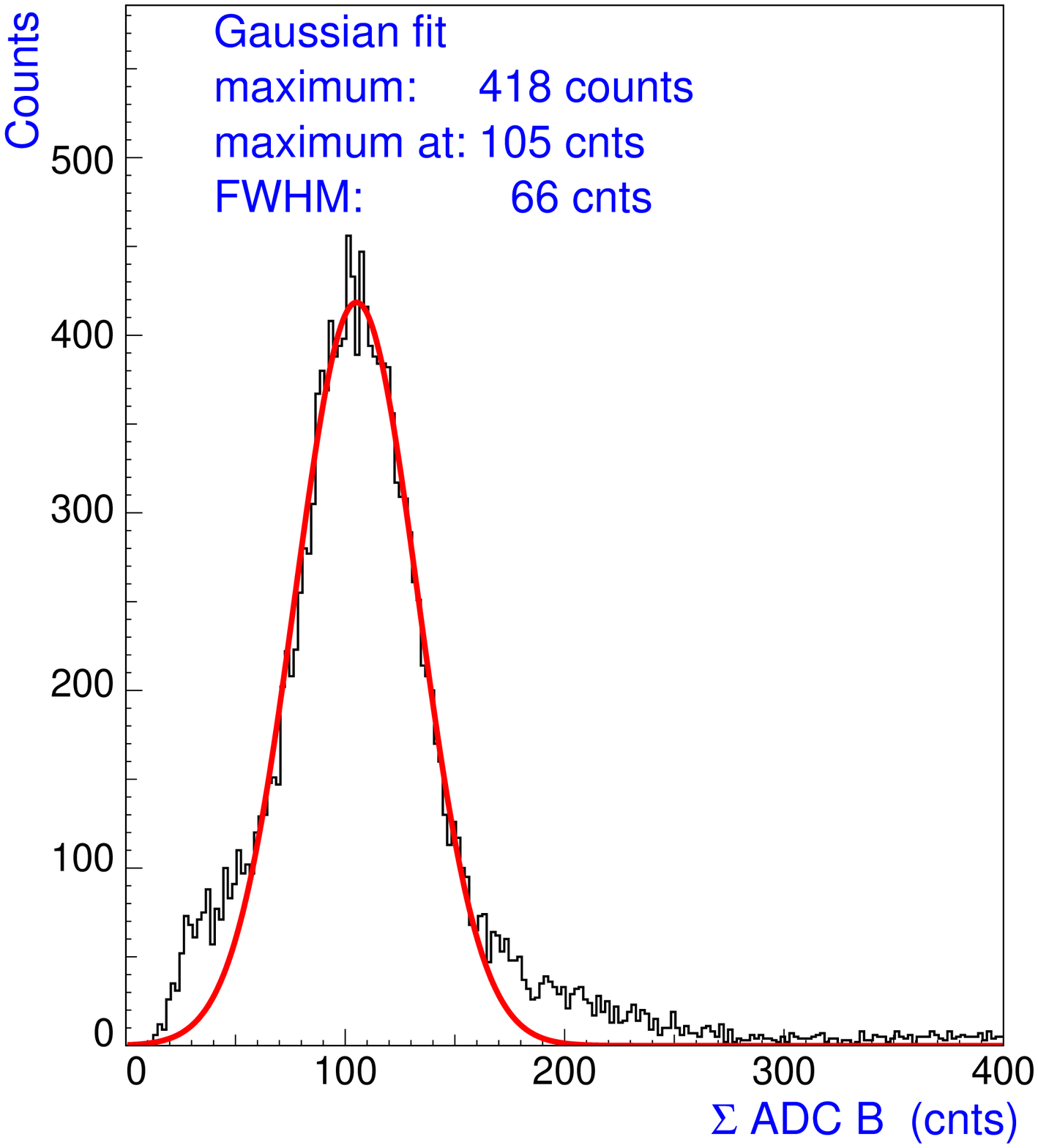}
  \includegraphics[width=0.6\textwidth]{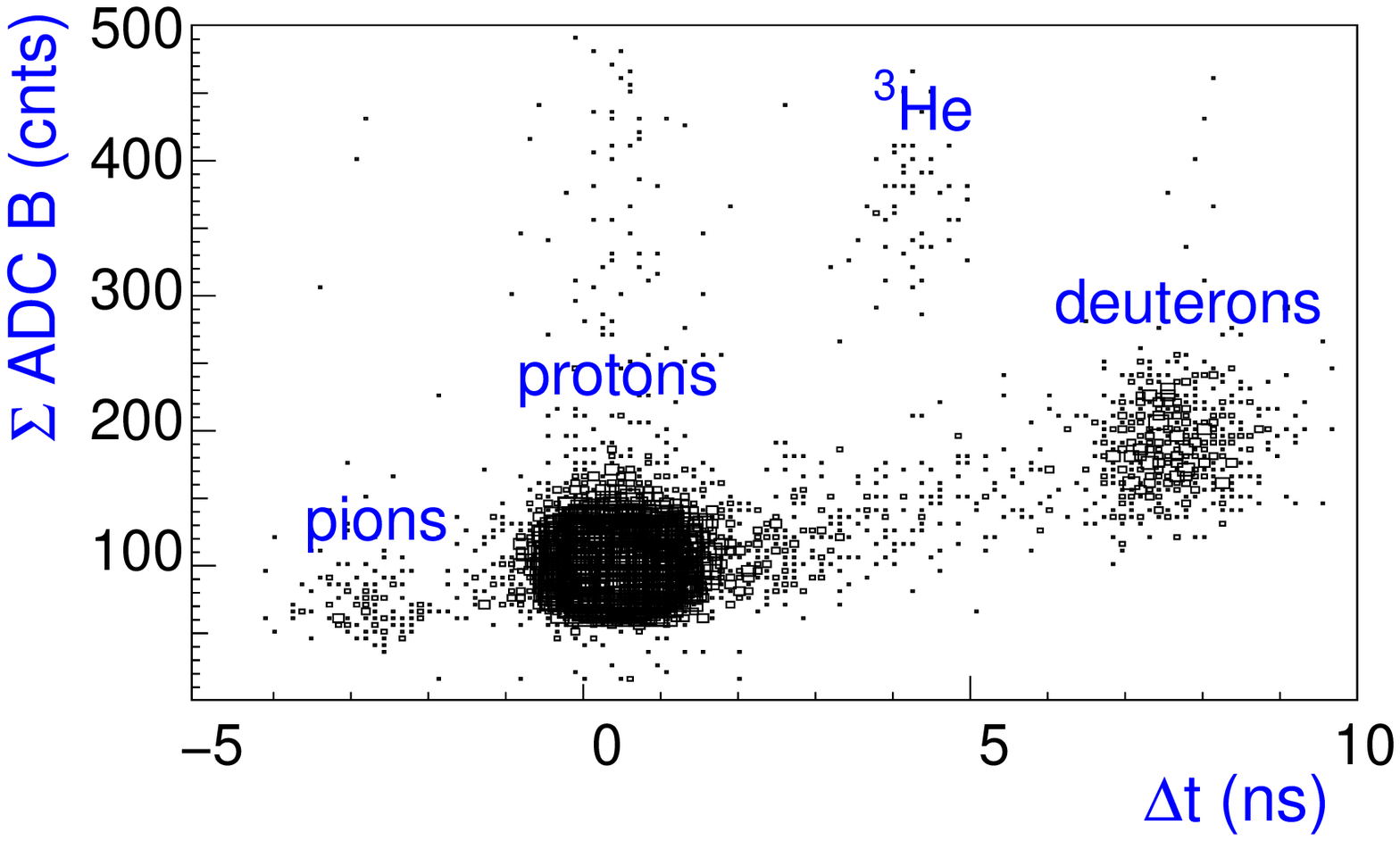}
  \caption{Distributions of the pulse height sum over all channels of
    one detector plane for the beam of different particle species
    (top). From the Gaussian fit to the proton peak a relative energy
    resolution $\Delta E/E\sim$ 60\,\% was derived.  The pulse height
    sum is shown vs.\ the time-of-flight so that $\pi^+$, proton,
    deuteron, and $^3$He were separated (bottom).}
  \label{fig:GSIcb-particleID}
\end{figure}

The energy response of the fibre detector was studied in the beam of
different particle species. Fig.~\ref{fig:GSIcb-particleID}\,(top)
shows the distribution of the pulse height sum over neighbouring
channels of one detection plane.  From the Gaussian fit to the data a
relative variation in the measured energy deposition, $\Delta E/E$, of
60\,\% was derived for the dominant particle species.
Fig.~\ref{fig:GSIcb-particleID}\,(bottom) shows the energy loss vs.\
relative time-of-flight, in which dominant protons and subdominant
$\pi^+$, deuteron, and $^3$He were separated.

\section{Concluding remarks}
The performance of scintillating fibre detectors with MaPMTs was
extensively tested using electrons at the spectrometer facility at
MAMI, $^{12}${C} ions of 2\,$A$\/GeV energy as well as p$/\pi^+\!/$d
particles at GSI.  The hit position was reconstructed by calculating
the centroids of the charges collected from each read-out channel.
Good spatial accuracy and time resolution were obtained at practically
unity detection efficiency. The energy response to different particle
species was studied.

During tests at MAMI and at GSI the optical cross-talk caused by the
finite thickness of the PMT entrance window could be verified.
Indeed, this behaviour of the PMT has been also recognised by
Hamamatsu. Only very recently, a 32-channel PMT with black shielding
lamellae embedded in the glass window became commercially available
with significantly reduced cross-talk.

It is currently planned to perform in 2009 a first HypHI experiment at
GSI using three arrays of scintillating fibres as well as a first
\kaos experiment on the electro-production of hypernuclei at MAMI with
fibre detectors in the spectrometer's electron arm.
   
\section*{Acknowledgements}
Work supported in part by Bundesministerium f{\"u}r Bildung und
Forschung (bmb+f) under contract no.\ 06MZ176. T.R.~Saito and his
research group are granted by the Helmhotz Association and GSI as
Helmholtz-University Young Investigators Group VH-NG-239 and DFG
research grant SA 1696-1/1.

%%%%%%%%%%%%%%%%%%%%%%%%%%%%%%%%%%%%%%%%%%%%%%%%%%%%%%%%%%%%%%%%%%%%%
%                         BIBLIOGRAPHY                              %
%\bibliographystyle{mybibstyle}
%\bibliography{beamtests}
%%%%%%%%%%%%%%%%%%%%%%%%%%%%%%%%%%%%%%%%%%%%%%%%%%%%%%%%%%%%%%%%%%%%%

\clearpage

%%%%%%%%%%%%%%%%%%%%%%%%%%%%%%%%%%%%%%%%%%%%%%%%%%%%%%%%%%%%%%%%%%%%%
%                          FIGURES                                  %
%%%%%%%%%%%%%%%%%%%%%%%%%%%%%%%%%%%%%%%%%%%%%%%%%%%%%%%%%%%%%%%%%%%%%

%
%

%%%%%%%%%%%%%%%%%%%%%%%%%%%%%%%%%%%%%%%%%%%%%%%%%%%%%%%%%%%%%%%%%%%%%
%                             END                                   %
%%%%%%%%%%%%%%%%%%%%%%%%%%%%%%%%%%%%%%%%%%%%%%%%%%%%%%%%%%%%%%%%%%%%%

\end{document}